\documentclass[%
reprint,
superscriptaddress,
amsmath,amssymb,
aps,
prd,
floatfix,
twocolumn
]{revtex4}

\usepackage{color}
\usepackage{graphicx}
\usepackage{dcolumn}
\usepackage{bm}
\usepackage{hyperref}
\usepackage[normalem]{ulem}

\hypersetup{
    colorlinks=true,
    linkcolor=blue,
    citecolor=blue,
    filecolor=black,
    urlcolor=blue,
}

\renewcommand{\eqref}[1]{Eq.~\ref{#1}}

\newcommand{\Erad}{\ensuremath{E_{\rm other}}}

\begin{document}

\title{Estimating gravitational radiation from super-emitting compact binary systems}

\author{Chad Hanna}
  \email{crh184@psu.edu}
  \affiliation{The Pennsylvania State University, University Park, PA 16802, USA}

\author{Matthew C. Johnson}
  \email{mjohnson@perimeterinstitute.ca}
  \affiliation{Department of Physics and Astronomy, York University, Toronto, Ontario, M3J 1P3, Canada}
  \affiliation{Perimeter Institute for Theoretical Physics, Waterloo, Ontario N2L 2Y5, Canada}

\author{Luis Lehner}
  \email{llehner@perimeterinstitute.ca}
  \affiliation{Perimeter Institute for Theoretical Physics, Waterloo, Ontario N2L 2Y5, Canada}

\date{\today}

\begin{abstract}
Binary black hole mergers are among the most violent events in the Universe,
leading to extreme  warping of spacetime and copious emission of gravitational
radiation.  Even though black holes are the most compact objects they are not
necessarily the most efficient emitters of gravitational radiation in binary
systems.  The final black hole resulting from a binary black hole merger
retains a significant fraction of the pre-merger orbital energy and angular
momentum. A non-vacuum system can in principle shed more of this energy 
than a black hole merger of equivalent mass.  We study
these super-emitters through a toy model that accounts for the possibility that
the merger creates a compact object that retains a long-lived time-varying
quadrupole moment.  This toy model can capture the merger of neutron stars,
but it can also be used to consider more exotic compact binaries.  We hope that
this toy model can serve as a guide to more rigorous numerical investigations
into these systems.
\end{abstract}

\maketitle


\section{Introduction}
The first detections of gravitational waves by the LIGO and Virgo
collaborations~\cite{PhysRevLett.116.061102,Abbott:2016nmj} have ushered in the
era of gravitational wave astronomy.  Both observed events, GW150914 and
GW151226, are consistent with the merger of two stellar-mass black holes.
Although the collaborations have confirmed that there were no detections of
systems containing matter during advanced LIGO's first observing run, they
anticipate making detections of systems containing neutron stars in the next
few years as advanced LIGO's sensitivity increases~\cite{abbott2016upper}.

Binary black holes are regarded as ideal sources of strong gravitational waves,
and the recent detections have clearly confirmed this expectation. It is also
customary to consider binary black holes giving rise to scenarios capable
of radiating most efficiently in gravitational waves. This expectation is
supported by the fact that intense gravitational fields and high speeds are
probed in the merger of black holes and thus the peak strain can be
correspondingly large. Although a large peak strain allows a great deal of
information to be gleaned about the final, non-linear, stages of merger, it
does not necessarily yield the most detectable source of gravitational waves
nor does it imply binary black holes would emit the most energy for a given
mass in quasi-circular encounters.

A gravitational wave detector is sensitive to the total energy Impingent on the
detector within its frequency sensitivity, making a merger that emits more
energy over longer timescales potentially more detectable than a black hole
merger of equivalent mass.  In a binary black hole merger, a significant
portion of the orbital angular momentum at coalescence is retained in the angular
momentum 
of the resulting Kerr black hole~\footnote{As evidenced by the ability to simple,
first principle models, to relatively accurately capture the final state of the
system~\cite{Buonanno:2007sv,Kesden:2009ds}}. A merger event of equivalent mass
that could shed more of its initial orbital binding energy in a merger could emit a
larger total energy in gravitational waves, and therefore be, in principle,
more detectable. 

One candidate for such a ``super-emitting" event is a binary neutron star
system which forms a long-lived neutron star as a result of the merger.  Such
a star spins down as angular momentum is radiated away in gravitational
radiation~\cite{Sekiguchi:2012uc,Lehner:2014asa,Baiotti:2016qnr}. This
radiative stage will come to an end when either the massive neutron star that
results from the merger stops rotating or its time-dependent quadrupole moment
vanishes. Such different types of behavior allowed by neutron stars is also
representative, at a qualitative level, to the type of phenomenology that other
more exotic objects might provide.  Whether a binary neutron star or another
more exotic compact binary system can in principle emit more total energy than
a black hole binary depends, as we shall discuss below, to leading order on the compaction $C = G
M/R$ of the merging objects with $M,R$ the mass and radius of the object
respectively and $G$ Newton's constant. For neutron stars, $C \sim 10^{-1}$
and, more generally, exotic compact objects can have a somewhat larger
compaction and could emit significantly more total energy than a black hole
merger of equivalent mass.
Of course, for sufficiently large individual masses
in binary neutron star systems, a prompt collapse to a black hole ensues\footnote{That is,
when the total mass of the newly formed star is $\gtrsim 1.2 M_{\rm max}$ where
$M_{\rm max}$ is the maximum allowed mass for a stable star with the corresponding
equation of state (rotation providing the additional support above this value).}. Since
the gravitational wave strain is proportional to the chirp mass, low masses have
a low chirp mass and thus the detectability horizon is lower. On the other hand,
possible massive exotic compact objects --of which there is not certainty on possible
mass bounds-- 
could have a farther horizon and the absence
of detection from such objects would significantly constrain their existence.

Assessing the amount of energy that
can be radiated and the characteristics of gravitational waves produced 
by generic compact binary mergers provides guidance into the observational
opportunities ahead. 
In this note, we investigate the gravitational radiation emitted in a simple
model binary system composed of spherically symmetric, nonrotating compact
objects, and assess the detectability of such events. While we expect this
model to be quite far from realistic systems, it can provide a rule of thumb
for the qualitative range of signatures that could be expected in principle. In
particular, rather than concentrating on the radiative properties of binaries
composed of specific exotic objects
(e.g.~\cite{Palenzuela:2007dm,Chirenti:2007mk}) our model accounts for
alternatives in a simplistic manner with the goal of extracting the broad
qualitative features that general compact systems in quasi-circular mergers can
yield.  For the sake of presentation, and departing slightly from the standard
convention, throughout our discussion, we will reserve the term ``compact
object'' to refer to objects which are not black holes --e.g.
neutron stars (NS) and exotic compact objects (ECO). Such objects do not have an event
horizon but give rise to strong gravitational fields in their vicinity.  Additionally
we will often employ the term ``merger'' to denote the full dynamics of the 
binary; i.e. inspiral, collision, post-merger  and final state stages.

\section{A model for binary super-emitters}
Before describing our simple model for non-vacuum binaries, recall the standard
binary black hole scenario.  The absolute bound on the energy radiated from a
binary black hole system comes from black hole thermodynamics.  In order for
the entropy of the final merged black hole to be nondecreasing, the area of the
final black hole must be at least as large as the area of the two initial black
holes~\cite{bekenstein1973black}.  Since area scales as $M^2$, the final mass,
$M_f$, obeys $M_1^2 + M_2^2 \leq M_f^2$ for non-rotating black holes, implying
for equal mass binaries that $ \sqrt{2}M \leq M_f \leq 2 M$.  This yields a
maximum gravitational radiation efficiency of 29\% of the binary mass.  In
order for the final black hole to have no rotation, and for this bound to be
satisfied, all of its angular momentum would need to be radiated away during
the merger, or somehow extracted via the interactions with additional fields
(e.g. via the Penrose process~\cite{penrose1971extraction}), in
which case it would no longer be a pure vacuum binary black hole spacetime.  Numerical
relativity predictions for non-rotating, equal-mass black holes indicate such cases
have a lower efficiency radiation loss at $\sim 4\%$ of the binary
mass~\cite{pollney2011high} scaling with $\eta^2$, where $\eta =
M_1M_2/(M_1+M_2)^2 \leq 0.25$, for a fixed total mass.  The predicted
gravitational radiation efficiency has been confirmed observationally with
GW150914 which radiated $\sim 5\%$ of its mass in gravitational
waves~\cite{PhysRevLett.116.061102}.  We will therefore consider $\sim 4--5\%$
to be the threshold beyond which an equal mass compact binary system, with
individual objects without intrinsic angular momentum, will be
said to radiate more than its binary black hole counterpart.

\begin{figure*}
\includegraphics[width=5.5in]{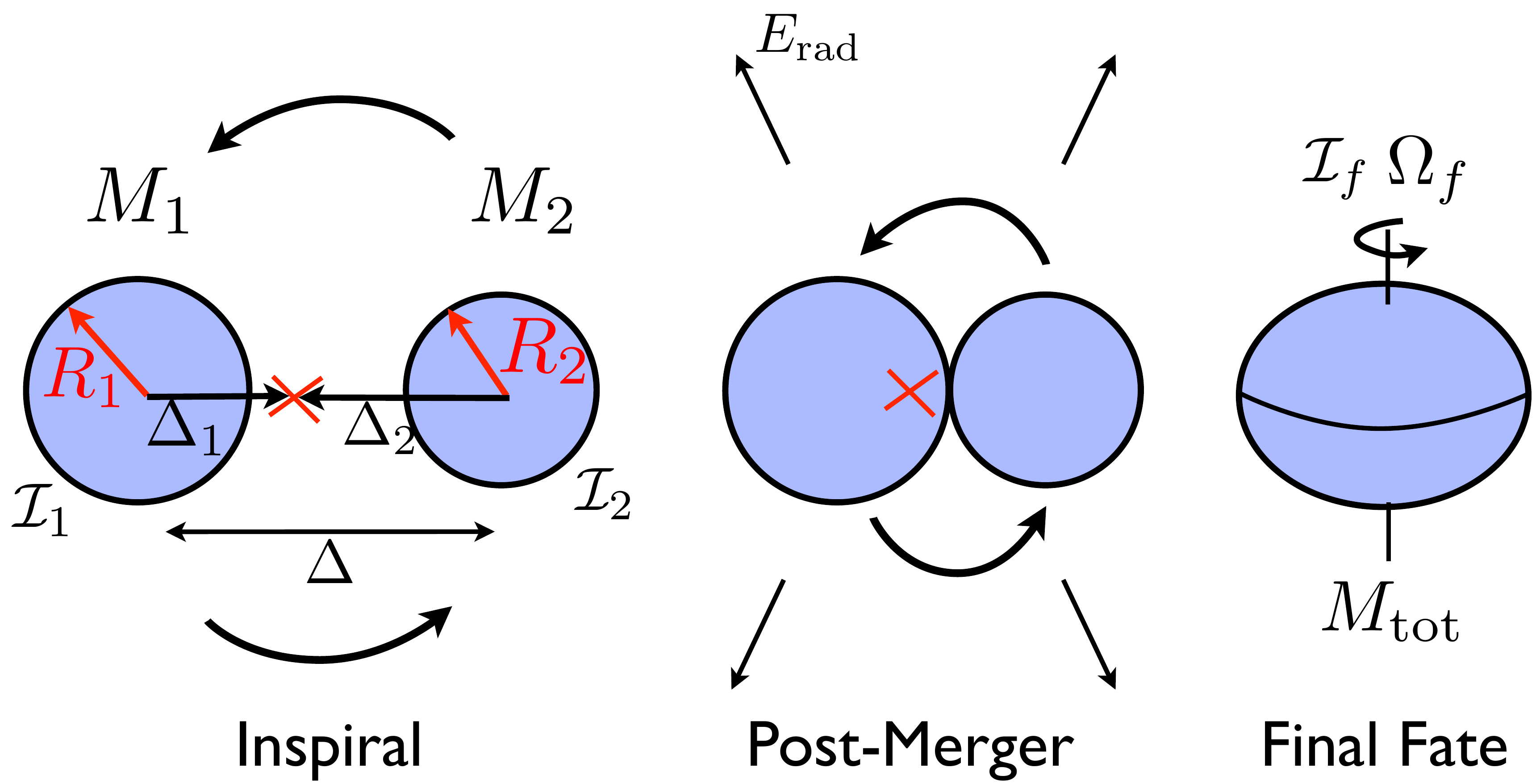}
\caption{
A toy model for compact object quasi-circular merger. During the inspiral phase, two objects
of mass $M_{1,2}$ and radius $R_{1,2}$ (we assume $M_1 \geq M_2$) undergo a
nearly spherical decaying Keplarian orbit about their center of mass (located a
distance $\Delta_{1,2}$ from masses 1 and 2 respectively). Once the objects
come into contact, they undergo a merged phase during which orbital energy is
converted to gravitational radiation and possibly other forms of radiation (of
total energy $\Erad$). After some time, the objects merge into an
axisymmetric object of constant density with mass $M_{\rm tot} \simeq M_1 +
M_2$ with moment of inertia $\mathcal{I}_f$ rotating at a constant angular
velocity $\Omega_f$.
}
\label{fig:toymodel}
\end{figure*}

Our model consists of two non-spinning, spherical compact objects as depicted
in Fig.~\ref{fig:toymodel}.  The heavier object has mass $M_1$, radius $R_1$,
and moment of inertia $\mathcal{I}_1$; the lighter object mass $M_2$, radius
$R_2$, and moment of inertia $\mathcal{I}_2$. Our toy model includes three
phases of evolution: inspiral, post-merger, and final fate. In the inspiral phase,
the persistent emission of gravitational radiation causes the compact objects
to undergo a slowly decaying nearly circular Keplerian orbit about the center
of mass. The center of mass is located a distance $\Delta_{1,2}$ from each
respective mass, and a distance $\Delta \equiv \Delta_1 + \Delta_2$ from each
other. As the objects come into contact, we define the post-merger phase as the time
at which the separation between the objects is approximately constant and some
fraction of the orbital energy is converted to gravitational radiation.  This
is of course a stark difference with respect to a binary black hole coalescence, which is
characterized by a quick plunge from an approximate inner most stable 
circular orbit (ISCO).  
We also allow for the possibility that energy
is released in other non-gravitational-wave forms e.g., electromagnetic
radiation, neutrinos and mass shedding denoted as $\Erad$. After some time, the
merged object achieves its ``final fate'' which is defined by having no
residual quadrupole moment (e.g. an axisymmetric or non-rotating object). The
mass of the final object is assumed to be approximately conserved, $M_{\rm tot}
\simeq M_1 + M_2$, i.e., the stars do not shed significant mass, and any
residual rotational energy is locked into the spin of the final object at
angular frequency $\Omega_f$. Below, it will be useful to write quantities in
terms of the compaction of the initial and final objects $C_{i} = G M_{i} /
R_{i}$; with $i=1,2,f$ labelling object $1$, $2$ or the final object
respectively. 

\subsection{Energy emitted in gravitational waves}

During the inspiral phase, the objects orbit at an instantaneous angular
frequency $\Omega_{\rm insp}$ about the center of mass, which to leading order, is
given by the Keplerian expression $\Omega_{\rm insp} = \sqrt{G M_{\rm tot} /
\Delta^3}$. The center of mass is located a distance $\Delta_{1,2} = M_{2,1}
\Delta / M_{\rm tot}$ from mass 1 and 2 respectively. Setting the energy to
zero for infinitely separated objects, the total change in energy during the
inspiral phase due to gravitational wave emission can be estimated as the total
energy at the moment when the objects come into contact,
\begin{align}\label{eq:insp}
\Delta E_{\rm insp} &= 
	\left(
	- \frac{G M_1 M_2}{\Delta}  
	+ \frac{1}{2} \mathcal{I} \Omega_{\rm insp}^2 
	\right)_{\Delta = R_1 + R_2}.
\end{align}
The moment of inertia $\mathcal{I}$ is given by,
\begin{align}
\mathcal{I} &= 
	\tilde{\mathcal{I}}_1 M_1 R_1^2 
	+ \tilde{\mathcal{I}}_2 M_2 R_2^2
	+ \mu \Delta^2,
\end{align}
where $\mu = M_1 M_2/M_{\rm tot}$ is the reduced mass and we have chosen to
parameterize the moment of inertia of the initial state objects by
dimensionless constants $\tilde{\mathcal{I}}_{1,2}$. For constant density
spheres, $\tilde{\mathcal{I}}_{1,2} = 2/5$, while a more realistic distribution
of mass would yield somewhat smaller values of $\tilde{\mathcal{I}}_{1,2}$
(e.g.~\cite{Grigorian1997}).

Once the objects come into contact, they no longer execute a Keplerian orbit
and the understanding of the binary's dynamics requires a more delicate
analysis including further physics. For instance, in the case of binary
neutron stars, non-linear, general
relativistic magnetohydrodynamics accounting for relevant microphysics
in the system (see e.g.~\cite{Lehner:2014asa} and
references therein) should be considered. Naturally, the physics ingredients
required for analysing relevant scenarios depend on the type of objects
being considered. With the purpose of deriving an upper-bound for generic
objects, here we 
envision a highly idealized post-merger phase, where at the initial contact stage,
the objects do not deform and simply fuse into a ``Janus dumbell'' as
illustrated in figure~\ref{fig:toymodel}. This phase terminates in what we
refer to as ``the final fate" where no further gravitational waves are emitted.
Naturally this phase either describes a non-rotating object or an axisymmetric,
stationary, object. We recognize that during this post-merger phase there might be
non-negligible energy loss via electromagnetic \& scalar radiation and particle emission,
(e.g. neutrinos in the case of neutron stars) which must also be accounted for (for
some recent examples see~\cite{Palenzuela:2015dqa,Foucart:2015gaa,East:2015vix,Sekiguchi:2016bjd,Lehner:2016lxy,Radice:2016dwd}). 
The change in energy of the binary system
due to gravitational wave emission between these two stages is given by the
difference between the orbital energy at the beginning of the merged phase and
the residual energy in the rotation of the final object as well as any
non-gravitational-wave energy loss, $\Erad$,
\begin{align}\label{eq:ewd}
\Delta E_{\rm pm} &=
	- \left( \frac{1}{2} \mathcal{I} \Omega_{\rm insp}^2 \right)_{\Delta = R_1 + R_2} 
	+ \frac{1}{2} \mathcal{I}_f \Omega_f^2 
	+ \Erad,
\end{align}
where the moment of inertia of the final merged object is $\mathcal{I}_f$.
    
Combining Eq.~(\ref{eq:insp}) and Eq.~(\ref{eq:ewd}), the total energy
emitted in gravitational radiation is given by the gravitational
potential energy of the two masses in contact less the residual rotational
energy of the final object,
\begin{align}
E_{\rm GW, CO} &=
	- \left(\Delta E_{\rm insp} + \Delta E_{\rm pm} \right) \nonumber \\
	       &=
	\frac{G M_1 M_2}{\Delta} 
	- \frac{1}{2} \mathcal{I}_f \Omega_f^2 
	-  \Erad \label{radiation1}.
\end{align}
Examining the relative contribution from the inspiral and post-merger phases, if
little energy is radiated or tied up in the rotation of the final state object,
then more gravitational radiation could be emitted during the post-merger phase than
during inspiral. For example, in the case of identical objects ($M_1 = M_2 =
M$, $R_1 = R_2 = R$, $C_1=C_2=C$, $\tilde{\mathcal{I}}_1=\tilde{\mathcal{I}}_2
= \tilde{\mathcal{I}}$) and $\Omega_f \rightarrow 0$ (i.e. a phase ending in a
non-rotating object), we have
\begin{align}\label{eq:upper_equalmass}
\max \left[ E_{\rm GW, CO} \right] &=
	\underbrace{\frac{(1-\tilde{\mathcal{I}})MC}{4}}_{\Delta E_{\rm insp}}
	+ \underbrace{\frac{(1+\tilde{\mathcal{I}})MC}{4}}_{\Delta E_{\rm merg}}
	= \frac{MC}{2}.
\end{align}
Note that for constant density spheres ($\tilde{\mathcal{I}} = 2/5$), $70 \%$
of the total energy in gravitational radiation can be emitted during the post-merger
phase. This stage in the evolution can therefore be very significant in
assessing the detectability of compact object mergers, as we discuss in more
detail below. 

How close a realistic scenario gets to the upper bound in
\eqref{eq:upper_equalmass} depends on the details of the merger scenario,
the composition of the merging objects and the possible final fate of the
merger. These will determine the amount of energy radiated and stored in the 
spin and mass of the final object. For ECOs, our knowledge is naturally restricted
to a few proposed models and further limited by the small number of works that have explored
the non-linear regime described by the merger; indeed, to our knowledge, only boson
stars have been studied in this context~\cite{Palenzuela:2007dm,Liebling:2012fv}.

The case of neutron star mergers is on better footing, with significant efforts exploring
their complex phenomenology during and after coalescence and including several mechanisms
through which the system can loose energy. A detailed estimate of the post-merger object's
behavior must account for mass shedding, emission in
electromagnetic radiation and neutrinos in $\Erad$, and the lifetime
of configurations with a time-dependent quadrupole to determine the rotational
energy of the final object. As a bound on the total energy emitted electromagnetically, we
can appeal to the estimated energy emitted in short gamma ray bursts. Such bursts are
thought to be driven by neutron star mergers, and the total energy emitted is of order $E_{\rm GRB} \sim 10^{51}
\ {\rm ergs} \sim 10^{-3} \ M_{\odot}$, far below the energy emitted in
gravitational radiation. An even larger amount of energy is carried away by
neutrinos, but this is still arguably smaller than that in gravitational
waves~\cite{Sekiguchi:2011zd,Palenzuela:2015dqa}.  Simulations of binary
neutron star mergers also indicate that tidal effects arise during the
inspiral phase~\cite{Rosswog:1998hy,Palenzuela:2015dqa,Martin:2015hxa,Sekiguchi:2016bjd,Radice:2016dwd,Lehner:2016lxy,Baiotti:2016qnr},
but they are small, especially for higher compaction 
cases (which, as we shall discuss, are the cases which could radiate more energy
than the analog black hole system). These estimates imply that $\Erad$ will be small compared
to the energy emitted in gravitational radiation. Finally, for the right
 masses and EoS describing the neutron stars, the time-scale for
the post-merger phase can be as long as $10-10^4 \ s$
\cite{Ravi:2014gxa,Kastaun:2016yaf}, encompassing potentially millions of
pre-merger orbital periods (see Sec.~\ref{sec:waveforms}), and allowing a
significant amount of the pre-merger orbital energy to be emitted as
gravitational radiation during the post-merger phase.

Assuming that $\Erad$ is negligible, the expression
\eqref{eq:upper_equalmass} provides a bound that depends on the compaction of
the merging objects and the rotational energy of the final object. 
Assuming most of the gravitational binding
energy is released as gravitational radiation and does not get locked in the rotational
energy of the final object
we can estimate an upper bound on the energy that could be emitted in a
non-vacuum compact binary merger. In general, the total energy emitted
increases with compaction. An upper bound on the compaction of spherically
symmetric objects is given by Buchdal's theorem to be $C \leq
4/9$~\cite{Misner:1974qy}, which sets the maximum possible energy emitted in
gravitational radiation during our envisioned merger. The emitted energy is
greatest when the compaction of both objects is saturated at $C_1=C_2=4/9$,
yielding the bound, 
\begin{equation}\label{eq:coupperbound}
E_{\rm GW, CO} \leq \frac{4 \mu}{9} .
\end{equation}
For equal mass, non-spinning objects, this bound implies that the total energy
emitted in gravitational radiation can reach up to $\sim 11\%$ of the total
rest mass of the system. Compare this value to that of the expected $\sim
4-5\%$ of the total mass for equal mass, non-spinning binary black hole systems
(e.g.~\cite{Hannam:2009hh, pollney2011high, PhysRevLett.116.061102}).  

A lower limit on gravitational radiation in the post-merger phase can be computed by
assuming that the final state, different from a black hole, is ``instantaneously'' produced conserving
angular momentum. This fixes $\Omega_f$ to be,
\begin{equation}
\Omega_f \simeq \frac{\left( \mathcal{I} \Omega_{\rm insp}\right)_{\Delta = R_1+R_2} }{\mathcal{I}_f},
\end{equation}
which implies that the energy change during the merger is
\begin{align}\label{eq:prompt}
\Delta E_{\rm pm} &=
	-\left( 1 - \frac{\mathcal{I}}{\mathcal{I}_f} \right) \left( \frac{1}{2} \mathcal{I} \Omega_{\rm insp}^2 \right)_{\Delta = R_1 + R_2} 
	+ \Erad .
\end{align}
$\Delta E_{\rm pm}$ can be zero in this scenario.  However, unless the
initial ($\mathcal{I}$) and final ($\mathcal{I}_f$) moments of inertia are the
same, energy must be dissipated in the form of gravitational radiation,
electromagnetic radiation, neutrinos, mass shedding, etc. If $\mathcal{I}$ is
significantly different than $\mathcal{I}_f$, the energy loss can be an
appreciable fraction of the total rest mass of the system. This is a simple
motivation for the importance of searching for electromagnetic and neutrino
counterparts to compact object mergers
(e.g.~\cite{Bloom:2009vx,Metzger:2011bv,Andersson:2013mrx}), which could
elucidate the energy and momentum balance more effectively than observing a
single channel.

In the absence of other radiation, $\Erad$, and assuming no mass shedding,  
conservation of energy implies in
this prompt-final-state scenario that $\mathcal{I} \leq \mathcal{I}_f$, which
imposes a restriction on the configuration of the final fate object
that can be produced in the merger. For a spherical final object of constant
density characterized by a compaction $C_f$, this bound translates into $C/C_f
\geq \sqrt{7/8}$, or equivalently, $R_f / R \geq \sqrt{7/2}$. The lower bound
for the emitted energy from the system is provided in this case by the inspiral phase,
which we estimated in eqn. (\ref{eq:upper_equalmass}).

\begin{figure}[h!]
\includegraphics[width=\columnwidth]{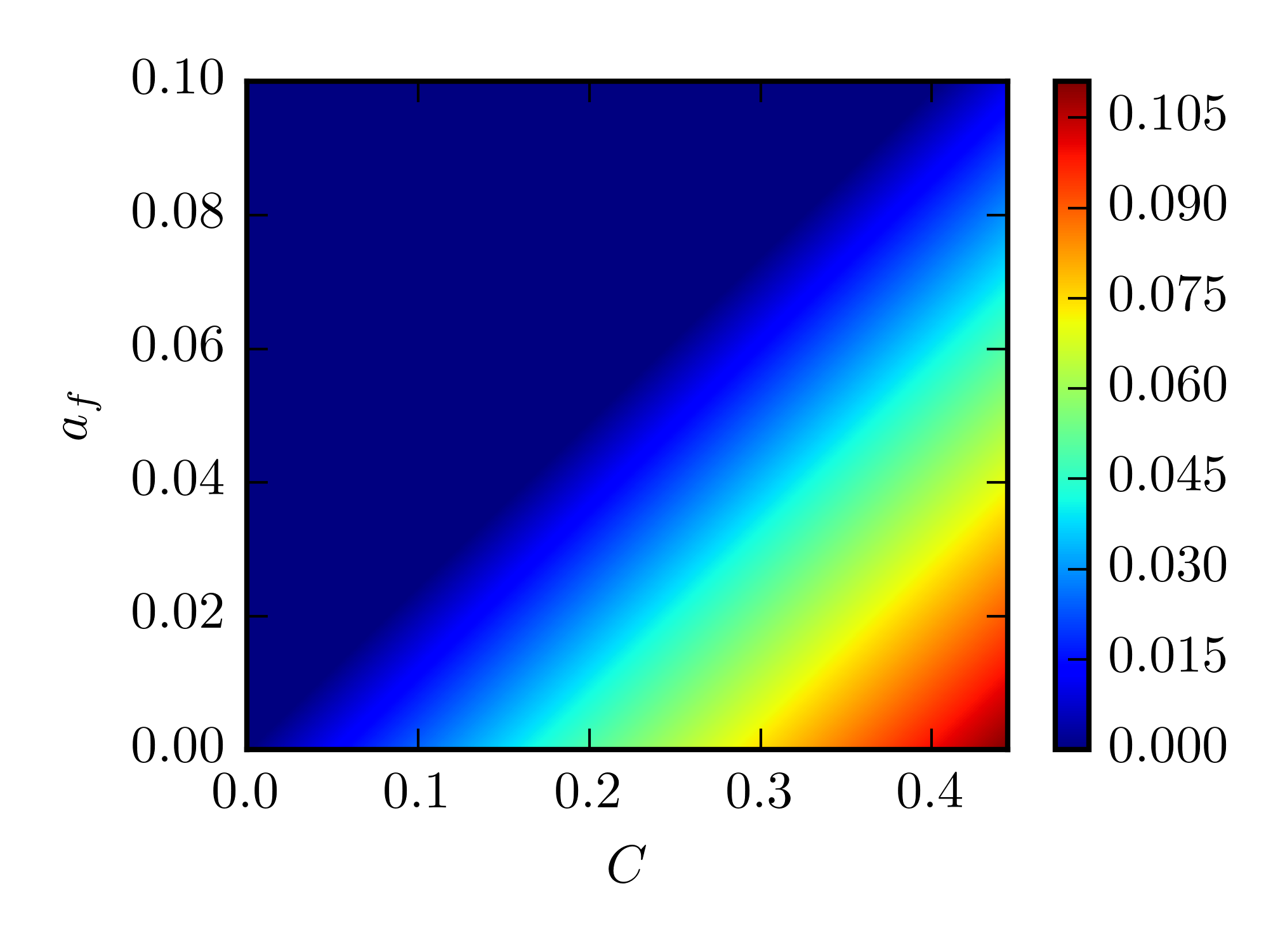}
\caption{The energy radiated per unit mass versus dimensionless final spin and final
compaction for equal object binaries described by \eqref{radiation1} and assuming
equal mass, equal radius, and equal compaction objects merge to form a final
object of maximal compaction (4/9) and uniform density ($\mathcal{I}_f =2/5$).
\label{fig:generalfigure}  
}
\end{figure}

Between these upper and lower limits, there will be a range of merger scenarios in which 
the energy emitted in gravitational waves is larger than the corresponding black hole system. 
In Fig.~\ref{fig:generalfigure} we show the energy radiated per unit mass versus dimensionless final spin and final
compaction for equal object binaries described by \eqref{radiation1}.  Assuming
equal mass, equal radius, and equal compaction objects merge to form a final
object of maximal compaction (4/9) and uniform density, the energy radiated can
exceed that of the equivalent binary black hole system ($\sim 4\%$) for a
family of final objects with dimensionless spin $a_f \equiv \Omega_f^2 / M^2
\lesssim 0.07$ and $C_f \gtrsim 0.15$.

\subsection{A more realistic scenario: neutron star mergers}

Numerical simulations of neutron star mergers provide a less idealized scenario
than the one described above. Such simulations are already furnishing a more complete
understanding of binary neutron star systems. However, such studies are computationally demanding due to
the intrinsic cost associated with each one, the larger parameter space describing
the binary (when compared
to binary black holes systems), and the inclusion of different physical effects that can play a role
over a disparate set of time and length scales. Here we employ some partial information inferred from simulations
to enrich our model at a modest level while maintaining its simplicity.
 We begin by considering the post-merger neutron star is characterized, to leading order, 
by its rotation frequency and moment of inertia. 
Assuming conservation of angular momentum during the collision, 
\begin{equation}
	\mathcal{I}_{f}  = \mathcal{I} \ \frac{ \Omega_{\rm insp} }{ \Omega_{\rm pm}},
\end{equation}
where $\mathcal{I}$ and $\Omega_{\rm insp}$ are evaluated at $\Delta =
R_1+R_2$.  During the post-merger phase, the change in energy due to emitted
gravitational radiation is 
\begin{align}
\Delta E_{\rm pm} &=
	- \frac{1}{2} \mathcal{I}_{f} \Omega_{\rm pm}^2  \\
	&= - \left( \frac{1}{2} \mathcal{I} \Omega_{\rm insp}^2 \right) \frac{\Omega_{\rm pm}}{\Omega_{\rm insp}}, \label{eq:ewd}
\end{align}
where we have omitted the terms in Eq.~\ref{eq:ewd} associated with other forms
of radiation and we assume that the final fate object does 
not rotate.

An estimate of the ratio of angular frequencies can be obtained in numerical
relativity simulations, which indicate that $\Omega_{\rm pm} / \Omega_{\rm
insp} \simeq 2$
(e.g.\cite{Kastaun:2014fna,Palenzuela:2015dqa,Foucart:2015gaa,Bauswein:2015yca,Hotokezaka:2016bzh})
is a reasonable expectation, yielding,
\begin{equation}
	\Delta E_{\rm pm} = - \mathcal{I} \Omega_{\rm insp}^2,
\end{equation}
and therefore
\begin{equation}
	E_{\rm GW, NS} = \frac{1}{2} \mathcal{I} \Omega_{\rm insp} + \frac{G M_1 M_2}{\Delta}.
\end{equation}
Note that this is larger than the binding energy of the individual objects, and
therefore in principle larger than in the toy model presented above. This is because:  (i)
we have neglected additional radiative degrees of freedom (which need not be significant
depending on the objects involved) and (ii) some of the internal
binding energy of the two objects has been converted to gravitational radiation
in deforming the merging compact objects into the final compact object, which
is reflected in the change in the moment of inertia. 

In the case of the merger between identical objects and assuming the absence of
other radiative degrees of freedom, we can again compare the total energy
emitted by a binary neutron star system to an equivalent mass binary black hole
system. The total energy emitted in the neutron star system is $E_{\rm GW, NS}
= 17 MC/20$. Comparing against the energy emitted in the equivalent black hole
system, the neutron star system will emit more total gravitational radiation as
long as $C > .14$. This is comparable to reasonable compactions in neutron
stars, implying that in principle a neutron star binary can emit more total
gravitational radiation than the equivalent black hole system if there is no
significant dissipation in other forms of energy and the final object is non-rotating;
i.e. for relatively low total mass binaries that avoid collapse to a black 
hole (e.g.~\cite{Bauswein:2013jpa,shibata2015numerical}).

\section{The waveforms}\label{sec:waveforms}
The two polarizations of gravitational wave in our model are given by
\begin{align}\label{eq:hplus}
	h_{+}      &= 	\frac{4G \Omega^2 \mu^2 \Delta^2}{r}
			\cos (2 \Omega t) \nonumber \\
	h_{\times} &= 	\frac{4G \Omega^2 \mu^2 \Delta^2}{r}
			\sin (2 \Omega t).
\end{align}
During the inspiral phase, both $\Delta$ and $\Omega$ evolve in time as the
orbit decays. During the merger phase, $\Delta = R_1 + R_2$ and only the
frequency changes in time. The inspiral is associated with a chirp, e.g. an
increase in frequency, but the merger is associated with an anti-chirp, a
decrease in frequency. 

We can determine the time dependence of the frequency by computing the power
emitted in gravitational waves and comparing to the time derivative of the
orbital energy. The power emitted in gravitational waves is
\begin{equation}
	\frac{dE}{dt} = 
		- \frac{G}{5} 
		\langle 
			\frac{d^3 {\bf J}}{dt^3}
				\cdot 
			\frac{d^3 {\bf J}}{dt^3} 
		\rangle_t,
\end{equation}
where $J_{ij}$ is the reduced quadrupole tensor. Assuming rotation along the
$z$-axis of Cartesian coordinates, this is given by
\begin{equation}
	{\bf J} = \frac{\mu \Delta^2}{2}
	\left(
		\begin{array}{ccc}
			\cos (2 \Omega t) -\frac{1}{3} & \sin (2 \Omega t) & 0 \\
			\sin (2 \Omega t) & - \cos (2 \Omega t) -\frac{1}{3} & 0 \\
			0 & 0 & -\frac{2}{3}
		\end{array}
	\right),
\end{equation}
which yields
\begin{equation}\label{eq:power}
	\frac{dE}{dt} =
		- \frac{32 G \mu^2 \Delta^4 \Omega^6}{5}.
\end{equation}

During the inspiral phase, we can compare the power emitted in gravitational waves
Eq.~\ref{eq:power} to the time derivative of the orbital energy 
\begin{equation}
	\frac{dE}{dt} =
		\left[ 
			\frac{G M_1 M_2}{\Delta^2} 
			\frac{d\Delta}{d\Omega}
			+ \mathcal{I} \Omega
		\right] 
		\frac{d\Omega}{dt},
\end{equation}
in order to obtain the time derivative of the orbital frequency. During the
early stages of inspiral where $\Delta \gg R_1, R_2$, we obtain
\begin{equation}\label{eq:domegainsp}
	\frac{d \Omega}{dt} =
		96 G^{5/3} \mu M_{\rm tot}^{2/3} \Omega^{11/3}.
\end{equation}
During the merger phase, $d\Delta / dt = 0$ in our simple model, and we therefore have
\begin{equation}\label{eq:domegawd}
	\frac{d \Omega}{dt} = 
		- \frac{32 G \mu^2 \Delta^4 \Omega^5}{5 \mathcal{I}_{\Delta = R_1+R_2} }.
\end{equation}

Setting the time of contact between the two compact objects to be $t=0$, the
solutions to Eq.~\ref{eq:domegainsp} and~\ref{eq:domegawd} are
\begin{align}
	\Omega(t) &= 
		\frac{\Omega_{\rm insp}}{\left( 1 - \alpha_{\rm insp} t \right)^{3/8}}, 
			\ \ t < 0 \\
	\Omega(t) &= 
		\frac{\Omega_{\rm insp}}{\left( 1 + \alpha_{\rm pm} t \right)^{1/4}}, 
			\ \ t > 0,
\end{align}
where the time constants $\alpha_{\rm insp}$ and $\alpha_{\rm pm}$ are given by 
\begin{equation}
	\alpha_{\rm insp} =
		\frac{8 \Omega_{\rm insp}^{8/3}}{3} 
		96 G^{5/3} \mu M_{\rm tot}^{2/3} 
	= 
		\frac{32 C^4}{GM},
\end{equation}
and
\begin{equation}
	\alpha_{\rm pm} =
		4 \Omega_{\rm insp}^4 
		\frac{32 G \mu^2 \Delta^4 }{5 \mathcal{I}_{\Delta = R_1+R_2} }
	=
		\frac{16 C^4}{7 GM},
\end{equation}
where $\Omega_{\rm insp} = \sqrt{G M_{\rm tot}} (R_1+R_2)^{-3/2} =
C^{3/2}/(2GM)$ and in the second equality here and above we present the result
for identical merging objects.

\begin{figure}
\includegraphics[width=3.in]{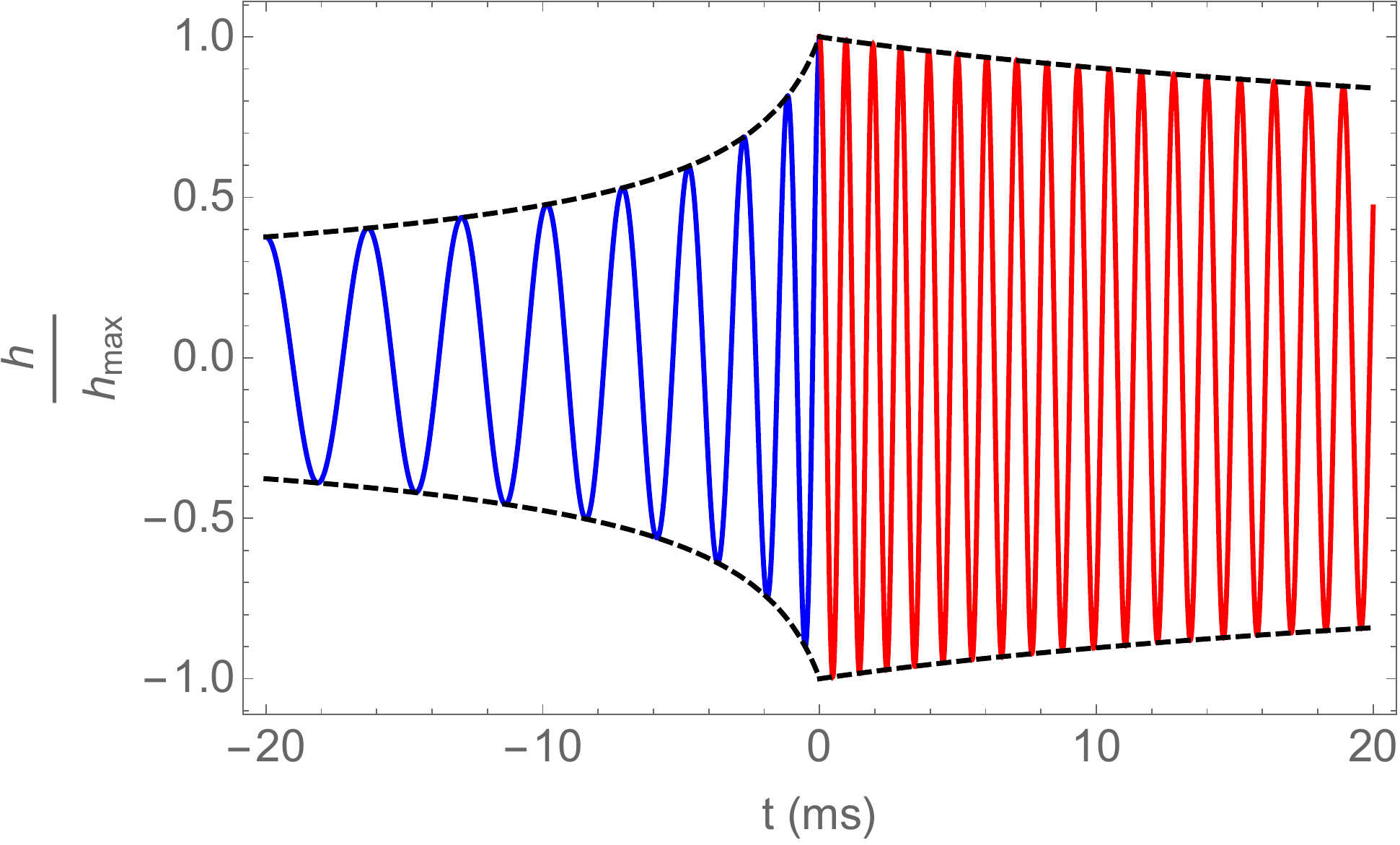}
\caption{
An example of the waveform Eq.~\ref{eq:hplus} for the merger of compact objects
including an inspiral phase (blue) and ring down phase (red) for identical
objects of mass $M = M_{\rm \odot}$ with $C=.1$. The chirp during inspiral has
a higher pitch than the anti-chirp during ring down, creating a
quasi-monochromatic signal during the post-merger phase.
}
\label{fig:waveformfigure}
\end{figure}

In Fig.~\ref{fig:waveformfigure} we sketch an example of the wave form for
identical objects with $C = .1$, including both the inspiral and post-merger phases.
The time constants set the characteristic rate of change of the frequency,
while $\Omega_{\rm insp}$ sets the characteristic frequency.  For identical
objects of mass $M = M_{\rm \odot}$ and compactions of order $C=0.1$, the time
constants are approximately $\alpha_{\rm insp}^{-1} \simeq 1.6 \ {\rm ms}$ and
$\alpha_{\rm pm}^{-1} \simeq 20 \ {\rm ms}$. This can be compared to the orbital
period at contact, $\Omega_{\rm insp}^{-1} = 0.3 {\rm ms}$, which is far
smaller than the decay time. Comparing with lifetimes of as long as $10-10^4 \
{\rm s}$ for the time dependent quadrupole of merged neutron
stars~\cite{Ravi:2014gxa,Kastaun:2016yaf}, we also see that the frequency can
decay appreciably in realistic systems, which implies that a significant amount
of energy is radiated in gravitational waves. More generally, the number of
cycles compared to the time constants scales like $\alpha_{\rm wd,
insp}/\Omega_{\rm insp} \sim C^{-5/2}$, which can be quite large for small
compaction. Therefore, during the post-merger phase, the comparatively large
frequency and slow decay of the frequency give rise to a nearly monochromatic
signal. 
 
Of course, our simple model captures the post-merger waveforms of realistic
systems such as neutron star mergers only at a qualitative level. Missing from
the model are: modulations in the waveforms resulting tidal effects as well as from
compression/decompression of the merged object (in the case of neutron stars,
see
e.g.~\cite{Takami:2014tva,Bauswein:2015yca,Bernuzzi:2015rla,Lehner:2016lxy}),
the main frequency of the early post-merger stages differing from $\Omega_{\rm
insp}$ by a factor of $\simeq 2$ (for neutron stars, see
e.g.~\cite{Lehner:2016lxy}), additional modes resulting from normal modes of
the star, new modes resulting from possible
instabilities~\cite{East:2015vix,Radice:2016gym,Lehner:2016wjg}, as well as
additional physical  effects driven by angular momentum transport and cooling
(in neutron stars see,
e.g.~\cite{Anderson:2008zp,Sekiguchi:2012uc,Palenzuela:2015dqa,Foucart:2015gaa})
or interactions of characteristics fields of exotic compact objects (for the
case of boson stars see~\cite{Palenzuela:2007dm}). The relevance and importance
of each of these effects depends upon the nature of the merging objects. It is
possible to enrich the model in order to account for some of these
features, as has been done for neutron star mergers in
Ref.~\cite{Takami:2014tva}. However, careful modelling of specific systems is
beyond the scope of the present work, which is intended only to provide a
general rule of thumb for generic compact object mergers (detached as much as possible
from specific cases), and a benchmark for comparison with black hole
mergers of equivalent mass.

\section{Detectability}
\begin{figure*}
\includegraphics[width=3.in]{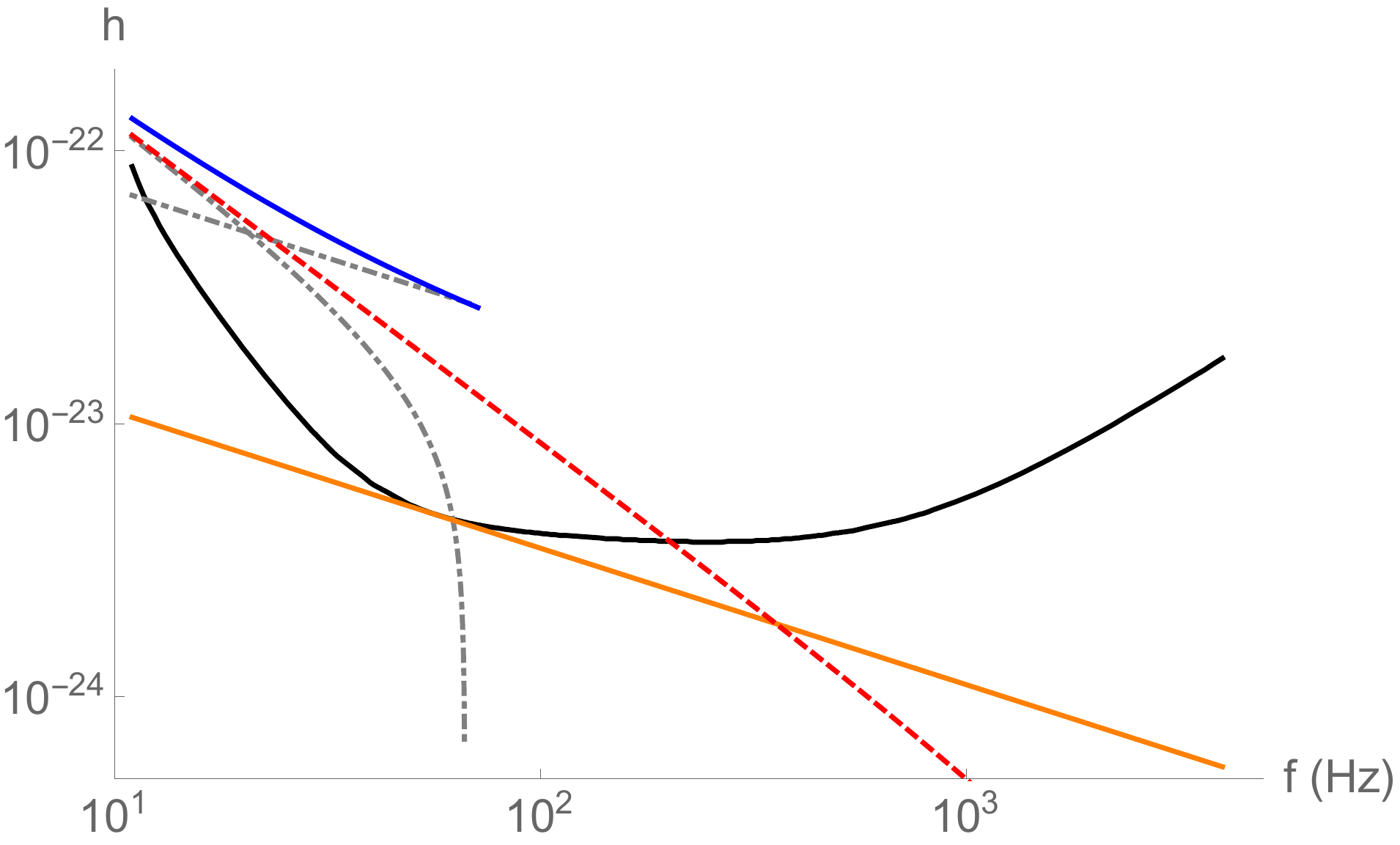}
\includegraphics[width=3.in]{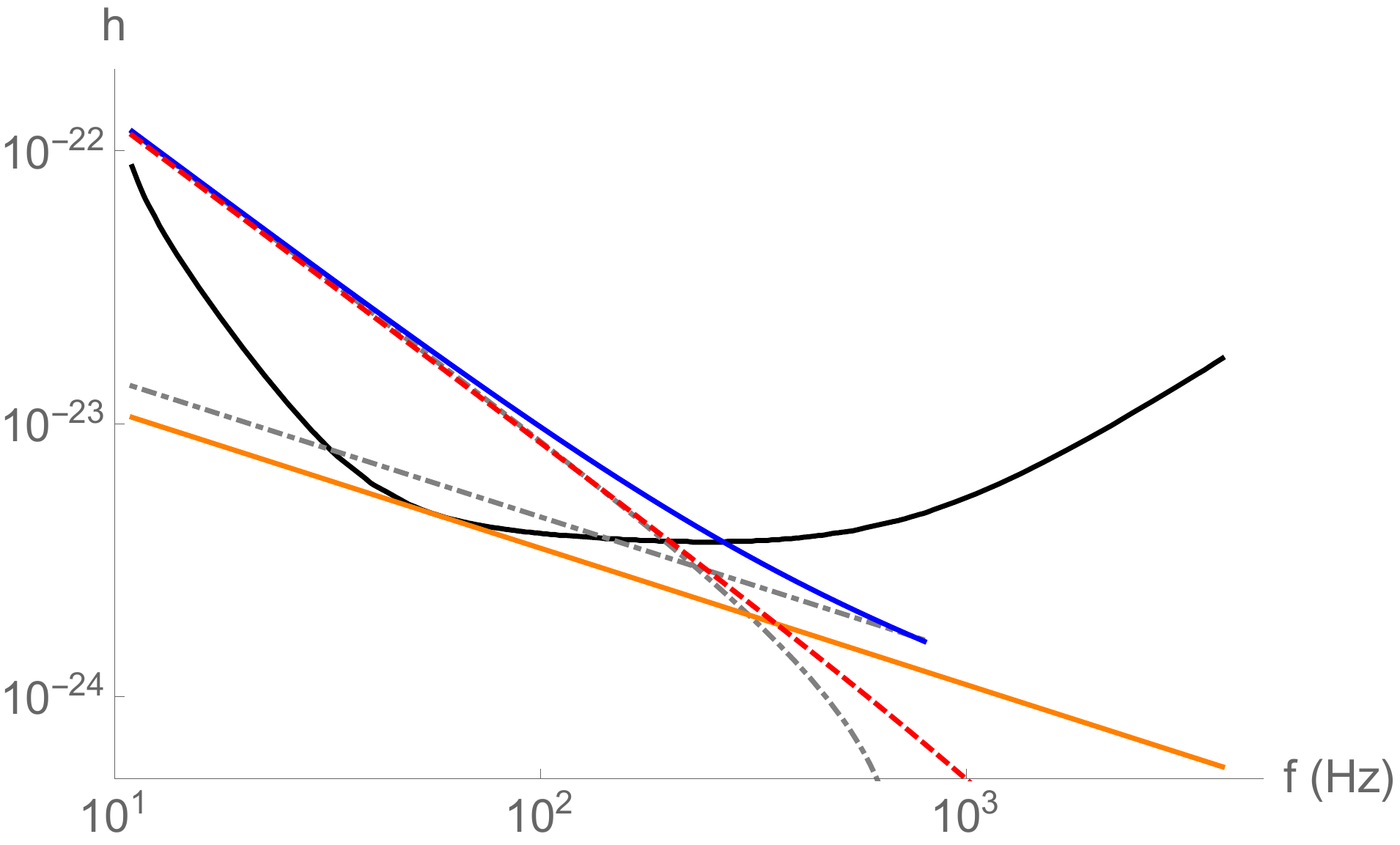}
\caption{Characteristic strain versus frequency for a variety of noise
scenarios and merger events. The black solid line is the projected
characteristic strain noise $|h_{\rm n}(f)|$ for advanced LIGO. The orange
solid line is the characteristic strain noise in a scenario with scale-invariant
sensitivity across the advanced LIGO frequency range. The red dashed line is the
characteristic strain of a black hole-black hole merger with $M_1 = M_2 = 1.25
M_{\rm \odot}$ at a redshift of $z=0.01$, computed in the PhenomD
phenomenological model. The grey dot-dashed lines are the characteristic strain
produced in the inspiral (Eq.~\ref{eq:charinsp}) and post-merger
(Eq.~\ref{eq:charm}) phases of a compact object merger in our toy model with
$M_1 = M_2 = 1.25 M_{\rm \odot}$ at a redshift of $z=0.01$ (the same as the
fiducial black hole system) with $\tilde{\mathcal{I}}_1 = \tilde{\mathcal{I}}_2
= 2/5$ and compaction $C=0.02$ (left panel) and $C=0.1$ (right panel). The
total characteristic strain for the compact object merger (Eq.~\ref{eq:charCO})
is shown as the solid blue line.}
\label{fig:sensitivity}
\end{figure*}

We have shown above that it is in principle possible for more total energy in
gravitational radiation to be emitted in the merger of compact objects than in
the equivalent mass black hole system. However, because gravitational wave
detectors are sensitive only over a range of frequencies, this extra
energy may or may not be easily detectable. In this section, we address the
detectability of the gravitational waves emitted by the model described above. 

An estimate of the signal to noise (SNR) can be obtained assuming an optimal
filter is applied to hypothetical time stream data. The square
of the SNR is given in frequency space by~\cite{allen2012findchirp} 
\begin{equation}\label{eq:SN}
\langle \, \text{SNR}^2 \, \rangle
	= 4 \int_0^{\infty} \frac{|h_{\rm char} (f)|^2}{S_n (f)} \,df,
\end{equation}
where $S_n(f)$ is the one-sided noise power spectral density.  We consider two
representative spectral densities. The first is the projected sensitivity of
advanced LIGO~\cite{Aasi:2013wya}, shown as the black line in
Fig~\ref{fig:sensitivity}, which for convenience we take to be infinity outside
the interval $10 \ {\rm Hz} < f < 4 \times 10^3 \ {\rm Hz}$.  The second is a scale 
invariant noise, defined in the same frequency interval as the LIGO sensitivity
curve, given by $h_{\rm n}(f) \simeq 3.5 \times 10^{-23} f^{-1/2}$. 

The one-sided signal power spectrum is defined as
\begin{equation}\label{eq:hchar}
h_{\rm char} (f)^2 
	\equiv \frac{5 G (1+z)^2}{8 \pi^2 D_L(z)^2} 
		f^{-2} \left| \frac{dE}{df}\right|_{(1+z) f} ,
\end{equation}
where $z$ is the redshift to the binary, $D_L(z)$ is the luminosity distance,
and $dE/df$ is the energy loss as a function of frequency evaluated at the
redshifted frequency. 

For a compact object merger with our model, the energy will vary differently
with frequency during the inspiral and post-merger phases, so we analyze these two
cases separately. Beginning with the inspiral phase, and identifying $f = \Omega
/ \pi$, we obtain
\begin{align}
\left| \frac{dE}{df} \right|_{\rm insp} \hspace{-15pt} (f < f_{\rm pm}) 
	&= \frac{\pi^{2/3} G^{2/3}}{3} \mu M_{\rm tot}^{2/3} f^{-1/3} \\
	&\hspace{12pt} - \pi^2\left(
		\tilde{\mathcal{I}}_1 M_1 R_1^2 + \tilde{\mathcal{I}}_2 M_2 R_2^2 
		\right) f , \nonumber
\end{align}
where we have used the Keplerian relation between $\Delta$ and $\Omega$, and
$f_{\rm pm}$ is the frequency at the end of the inspiral phase, given by $f_{\rm
pm} \equiv \sqrt{G M_{\rm tot} / \pi^2 } (R_1+R_2)^{-3/2}$. Comparing this to
the plunge frequency~\footnote{The transition from inspiral to the merger 
phase happens at the orbital radius past which there are no stable circular orbits, 
corresponding to the plunge frequency in the wave form.  To a rather good
approximation the plunge frequency, $\Delta_{\rm plunge}$, can be obtained via
first-principles as described in~\cite{Buonanno:2007sv,Kesden:2009ds}. For
nearly equal mass non-spinning black holes $\Delta_{\rm plunge} \approx 3.5 G
M_{\rm tot}$, yielding $f_{\rm pl} \simeq 0.048 (G M_{\rm tot})^{-1}$.}, for identical merging objects we have $f_{\rm pm} =
f_{\rm pl} C^{3/2} (\Delta_{\rm plunge}/G M_{\rm tot})^{3/2} \simeq 6.55 \
C^{3/2} f_{\rm pl}$. Note that for compactions larger than $C > .29$, the
merger frequency will be larger than the plunge frequency. During the post-merger
phase, we obtain
\begin{align}
\left| \frac{dE}{df} \right|_{\rm pm} \hspace{-10pt} ( f_{\rm ff} < f < f_{\rm pm} ) 
	&=& \pi^2 \mathcal{I}_{\Delta = R_1+R_2} f
\end{align}
where $f_{\rm ff}$ is the frequency at which the post-merger phase terminates.
Notice this frequency is $< f_{\rm merg}$ as a result of the spin-down of the
post-merger object due to the emission of gravitational waves; in particular if the
object retains its quadrupole, then $f_{\rm ff}=0$.

We consider the contributions to the characteristic strain from the inspiral
and post-merger phases separately:
\begin{equation}\label{eq:charCO}
h_{\rm CO} (f)^2 = h_{\rm CO, insp} (f)^2 + h_{\rm CO, pm} (f)^2
\end{equation}
where it is understood that $h_{\rm CO, insp}$ is defined in the interval
$f<f_{\rm merg}$ and $h_{\rm CO, merg}$ is defined in the interval $f_{\rm ff}<
f<f_{\rm merg}$. For the inspiral phase, we obtain:
\begin{align}\label{eq:charinsp}
h_{\rm CO, insp} (f)^2 
	&\simeq \frac{5}{8} (1+z)^{5/3}
		\left( \frac{{\rm Mpc}}{D_L(z)} \right)^2 \frac{\mu}{M_{\odot}} 
		\left(\frac{M_{\rm tot}}{M_{\odot}}\right)^{2/3} \nonumber\\
	&\hspace{12pt} \times 10^{-38} \ {\rm Hz}^{1/3} f^{-7/3} \nonumber \\
	&\hspace{12pt} - 0.78 (1+z)^{3}  
		\left[
			\frac{\tilde{\mathcal{I}_1}}{C_1^2} 
			\left( \frac{M_1}{M_{\odot}} \right)^3 
			+ \frac{\tilde{\mathcal{I}_2}}{C_2^2}
			\left( \frac{M_2}{M_{\odot}} \right)^3
		\right] \nonumber \\ 
	&\hspace{12pt} \times \left( \frac{{\rm Mpc}}{D_L(z)} \right)^2 
		f^{-1} \times 10^{-44} \ {\rm Hz}^{-1} .
\end{align}

For the post-merger phase, we obtain:
\begin{align}\label{eq:charm}
h_{\rm CO, pm} (f)^2
	&\simeq 0.78(1+z)^{3} \times
		\left( \frac{{\rm Mpc}}{D_L(z)} \right)^2
		f^{-1} \times 10^{-44} \ {\rm Hz}^{-1} \nonumber \\
	&\hspace{12pt} \times \left[ \frac{\tilde{\mathcal{I}_1}}{C_1^2} \left( \frac{M_1}{M_{\odot}}\right)^3 +  \frac{\tilde{\mathcal{I}_2}}{C_2^2} \left( \frac{M_2}{M_{\odot}}\right)^3 \right. \nonumber \\ 
	&\hspace{30pt} + \left. \frac{\mu}{M_\odot} \left( \frac{M_1}{M_\odot C_1} +  \frac{M_2}{M_\odot C_2} \right)^2 \right] .
\end{align}

We can now compare our compact binary model with a binary black hole waveform.
To properly capture the merger and ring-down phases, we adopt the ``PhenomD"
model~\cite{Khan:2015jqa}, one of the phenomenological waveform models that has
been tuned with numerical relativity simulations. 

In Fig.~\ref{fig:sensitivity}, we show the characteristic strain for a system
of two compact objects of mass $M_1 = M_2 = 1.25 M_{\rm \odot}$ at a redshift
of $z=0.01$ (the same as the fiducial black hole system) with
$\tilde{\mathcal{I}}_1 = \tilde{\mathcal{I}}_2 = 2/5$ and compaction $C=0.02$
(left panel) and $C=0.1$ (right panel). For these masses, a long-lived
neutron star is a likely outcome of the merger. Here, we have assumed that
$f_{\rm ff}$ lies outside the frequency range of the sensitivity curves. The
dashed grey curves show the contribution from $h_{\rm CO, pm}^2$ and $h_{\rm CO,
insp}^2$, while the blue curve shows their sum.  The relative SNR between the
compact object and black hole systems is both a function of the compaction and
the the strain noise.

\begin{figure}
\includegraphics[width=3.in]{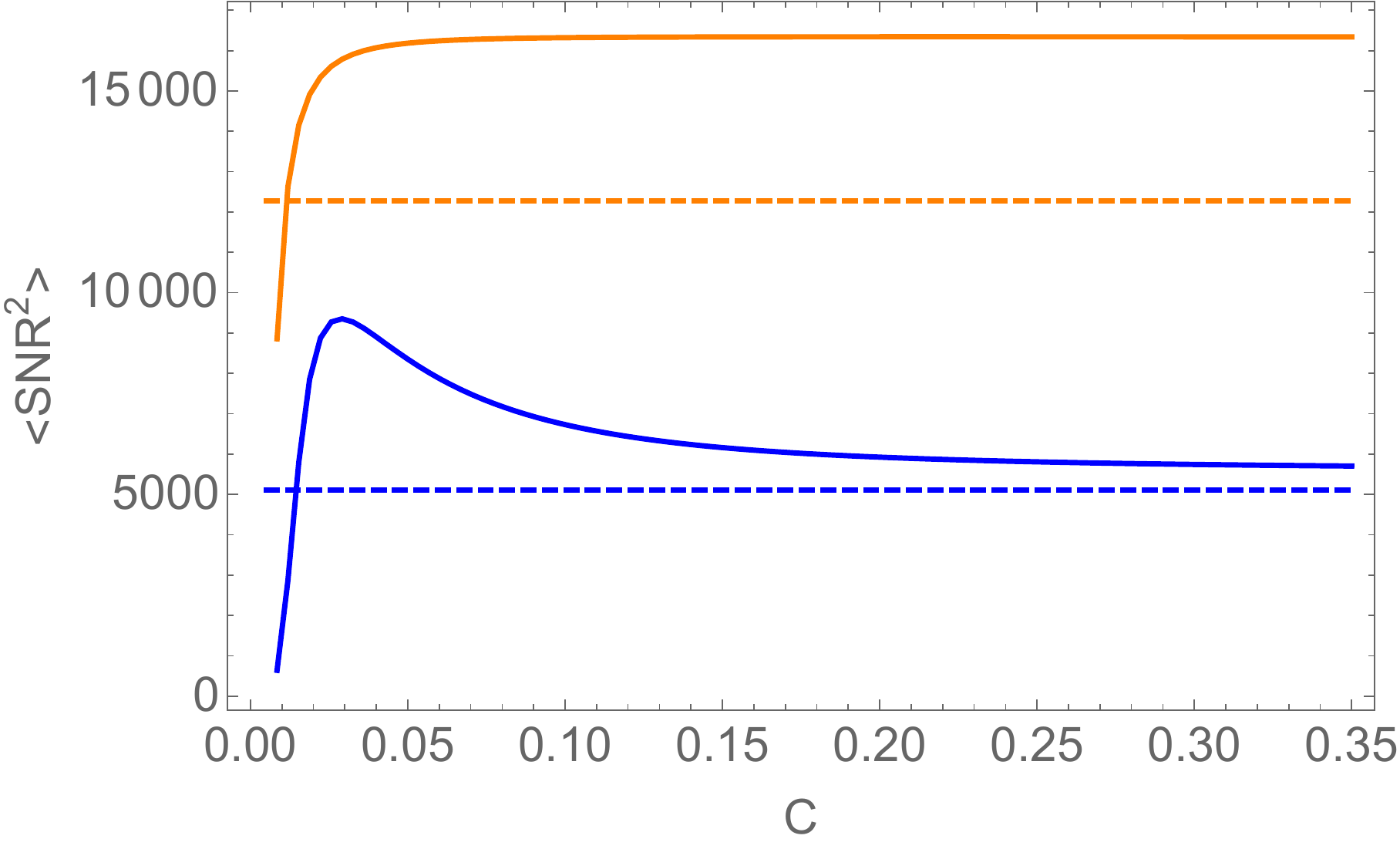}
\caption{The SNR squared (Eq.~\ref{eq:SN}) for the fiducial compact object
mergers shown in Fig.~\ref{fig:sensitivity} as a function of compaction for the
advanced LIGO (blue) and scale invariant sensitivity (orange) strain noise scenarios. The
SNR squared for the corresponding black hole systems are shown as dashed
lines.}
\label{fig:SNR}
\end{figure}

To examine the relative SNR quantitatively, in Fig.~\ref{fig:SNR} we plot the
SNR of the fiducial black hole system and an equivalent mass system of
identical compact objects from our model, as a function of the compaction. The
result for the advanced LIGO strain noise is shown in blue, with the
corresponding SNR for the black hole shown as the dashed blue curve.  The
result for the scale invariant strain noise is shown in orange, with the corresponding SNR
for the black hole shown as the dashed orange curve. We have again assumed that
$f_{\rm ff}$ lies outside (and below) the frequency range of the sensitivity
curves. The SNR is larger than the corresponding black hole system for
compactions larger than $C\sim.015$ for the advanced LIGO strain noise scenario
and $C\sim.01$ for the scale invariant noise scenario. The growth in the SNR with
compaction for the scale invariant sensitivity curve is in accord with the intuition that
systems where more total energy is emitted are also more detectable. For the
LIGO sensitivity curve, the story is a little more complicated. Since LIGO is
not sensitive to the majority of the gravitational radiation emitted in the
post-merger phase for this fiducial example at large compaction, the SNR first rises
then falls.  However, for all but the largest compactions where the post-merger
phase for the compact objects exits the LIGO sensitivity window, the cases
where more total energy is emitted are also more detectable.

Scanning a range of masses~\footnote{Naturally, this range of masses is too
large for NSs which would be bounded by $< 3 M_{\odot}$. However, mass bounds
for exotic compact objects are, so far, largely unconstrained.} between
$M_{\odot} < M < 30 M_{\odot}$, we found the compaction that yields the maximum
boost in SNR over the equivalent mass black hole system for the
advanced LIGO and scale invariant sensitivity strain noise scenarios.  For the advanced
LIGO case, the maximum boost in squared SNR is nearly flat over redshift and mass,
equal to $\sim 1.8-2.5$. The compaction at which the maximum boost occurs
increases with mass and is relatively independent of redshift, ranging from $C
\sim .03$ for $M =M_{\odot} $ to $C \sim .25$ for $M=30 M_{\odot}$. For the
scale invariant sensitivity scenario, the maximum boost is similar, relatively flat over
the parameter space and of order $\sim 1.3-2.0$. The compaction at which the
maximum occurs, however, is always saturated at the maximum value (see
Fig.~\ref{fig:SNR}), which in this case is $C = 4/9$.

We conclude that compact object mergers can emit more total gravitational
energy than their counterpart black hole system and can generally be more
detectable with present detectors. The boost in SNR is $\sim\sqrt{2}$.

\section{Conclusions}

In this note we have investigated a toy model binary merger of two compact objects that can in principle emit more energy in gravitational radiation than a black hole system of equivalent mass. While very simplistic, this model illustrates what might be possible in more realistic systems such as the merger of two neutron stars, or perhaps more exotic compact objects. We have found that the merger of objects with compactions of order $C \sim \mathcal{O}(0.1)$ and larger that avoid collapse to a black hole can yield more energy in gravitational radiation than the corresponding black hole system by up to a factor of roughly two. The SNR for such compact object mergers in a gravitational wave detector such as advanced LIGO can exceed that of the corresponding black hole merger by a factor of $\sim \sqrt{2}$ for somewhat smaller compactions $C \sim \mathcal{O}(0.03)$. This is in accord with the two-fold increase in total energy emitted in the most optimistic merger scenario.

Compact object mergers have the potential to radiate more energy and be more detectable than the merger of equivalent mass black holes, but do they?  This depends somewhat on how representative the assumptions underlying our toy model are. First, we have assumed that the only sink for orbital energy is in the form of gravitational radiation. This is clearly incorrect, as in any realistic scenario energy will be dissipated mechanically in the deformation of the objects, radiativelly through electromagnetic (and possibly scalar) radiation, or through neutrino (or other particles) emission. A complete treatment accounting for this would influence the total energy emitted, and the energy spectrum of the outgoing gravitational waves, which could change the prospects for detectability. We have also been generous in assuming that the post-merger phase terminates at frequencies lower than $10$Hz (i.e. below the lowest in the LIGO band). Relaxing this assumption would diminish the boost in detectability. Additionally, we have neglected general relativistic and tidal effects on the inspiral and post-merger phases. For large compactions as those we have considered, these effects are arguably small~\cite{Mora:2003wt,2014grav.book.....P}.
Finally, we have also considered non-spinning objects. Black holes could in principle be highly spinning and,
if relatively well aligned with the orbital angular momentum, their total radiated energy could be
considerably higher\footnote{Which could be as high as $12.5\%$ of the total mass of the system
for low eccentricity mergers. This estimate, obtained through~\cite{Buonanno:2007sv}, 
does not take into account the amount of energy radiated from the ISCO onwards, but this is small for 
this extreme scenario. Indeed,
the estimate is in excellent agreement with numerical results of highly spinning binary black hole mergers~\cite{Scheel:2014ina}.}. On the other hand, as in the case of neutron stars, a bound on the spin of non-vacuum compact objects could
exist to prevent mass shedding; consequently spin might only modestly boost the 
energy radiated in such cases. Nevertheless, according to current estimates of the projected spin along
the orbital angular momentum through gravitational waves, this is low~\cite{TheLIGOScientific:2016pea}; thus
our no-spin treatment in this work does not appear to be overly restrictive.

 These loose ends motivate a more systematic treatment which, in turn, involves specializing to specific objects (see e.g.~\cite{Takami:2014tva,Hinderer:2016eia})
with all the implications/physical requirements that their analysis would require.

Given the impending flood of data from LIGO and future gravitational wave detectors, we stand to learn a great deal about sources of gravitational radiation in our Universe. Since compact object mergers can in principle both yield more energy in gravitational radiation and be more detectable by advanced LIGO, it is important to keep an open mind about what surprises might await, and strive to gain some idea about how brig such a surprise might be.

\acknowledgments

MCJ and LL are supported by the National Science and Engineering Research
Council through a Discovery grant.  LL also thanks CIFAR for support. CH thanks
the NSF {\bf PHY-1454389}.  This research was supported in part by Perimeter
Institute for Theoretical Physics. Research at Perimeter Institute is supported
by the Government of Canada through the Department of Innovation, Science and
Economic Development Canada and by the Province of Ontario through the Ministry
of Research, Innovation and Science.

\bibliography{gravitywaves}

\end{document}